\documentclass[universe,article,accept,moreauthors,pdftex]{Definitions/mdpi}
\graphicspath{{Figures/}}
\usepackage{diagbox}
\usepackage{amsmath, amssymb}
\firstpage{1} 
\pubvolume{1}
\issuenum{1}
\articlenumber{0}
\pubyear{2024}
\copyrightyear{2023}
\externaleditor{Academic Editor: Firstname Lastname}
\datereceived{23 November 2023} 
\daterevised{15 December 2023} 
\dateaccepted{20 December 2023} 
\datepublished{ } 
\hreflink{Universe 2024, 10(1), 25} 
\doinum{10.3390/universe10010025}
\pdfoutput=1 

\usepackage{color}

\Title{Comparing numerical relativity and perturbation theory waveforms for a non-spinning equal-mass binary}
\TitleCitation{Title}

\Author{Tousif Islam $^{1,2,3\*}$\orcidA{},  Scott E. Field $^{1,3}$ and Gaurav Khanna $^{2,3,4}$}

\AuthorNames{Tousif Islam, Scott E. Field, Gaurav Khanna}
\AuthorCitation{Islam,T.;Field, S.; Khanna, G.}

\address{%
$^{1}$ \quad Department of Mathematics, University of Massachusetts, Dartmouth, MA 02747, USA\\
$^{2}$ \quad Department of Physics, University of Massachusetts, Dartmouth, MA 02747, USA\\
$^{3}$ \quad Center for Scientific Computing and Data Science Research, University of Massachusetts, Dartmouth, MA 02747, USA\\
$^{4}$ \quad Department of Physics and Center for Computational Research, University of Rhode Island, Kingston, RI 02881, USA}

\corres{Correspondence: tislam@umassd.edu}

\abstract{Past studies have empirically demonstrated a surprising agreement between gravitational waveforms computed using adiabatic-driven-inspiral point-particle black hole perturbation theory (ppBHPT) and numerical relativity (NR) following a straightforward calibration step, sometimes referred to as $\alpha$-$\beta$ scaling. Specifically focusing on the quadrupole mode, this calibration technique necessitates only two time-independent parameters to scale the overall amplitude and time coordinate. In this article, part of a special issue, we investigate this scaling for non-spinning binaries at the equal mass limit. Even without calibration, NR and ppBHPT waveforms exhibit an unexpected degree of similarity after accounting for different mass scale definitions. Post-calibration, good agreement between ppBHPT and NR waveforms extends nearly up to the point of the merger.  We also assess the breakdown of the time-independent assumption of the scaling parameters, shedding light on current limitations and suggesting potential generalizations for the $\alpha$-$\beta$ scaling technique.}


\begin{document}

\setcounter{section}{0} 
\section{Introduction}
Simulating binary black hole (BBH) mergers and understanding the morphology of the resultant gravitational waveforms enable us to develop tools for the rapid characterization of detected gravitational wave (GW) signals~\cite{Blackman:2015pia,Blackman:2017pcm,Blackman:2017dfb,Varma:2018mmi,Varma:2019csw,Islam:2021mha,bohe2017improved,cotesta2018enriching,cotesta2020frequency,pan2014inspiral,babak2017validating,husa2016frequency,khan2016frequency,london2018first,khan2019phenomenological}. An accurate description of BBH mergers requires numerically solving the Einstein equation for the two-body problem. This approach is known as numerical relativity (NR)~\cite{Mroue:2013xna,Boyle:2019kee,Healy:2017psd,Healy:2019jyf,Healy:2020vre,Healy:2022wdn,Jani:2016wkt,Hamilton:2023qkv} and is often computationally intensive, taking days to weeks to simulate a single BBH merger and its associated waveforms. In recent decades, significant advancements have been made in making NR codes more efficient in performing BBH simulations for systems where the masses of the two-component black holes are comparable, i.e., where $q:=\frac{m_1}{m_2} \lesssim 10$ is the mass ratio, $m_1$ is the mass of the primary black hole, and $m_2$ is the mass of the secondary black hole.
While NR simulations have begun to push beyond the $q=10$ barrier~\cite{yoo2022targeted,lousto2023study,Boyle:2019kee}, long-duration simulations with high accuracy remain challenging and parameter space explorations are mostly out of reach.

Alternatively, when $m_2 \ll m_1$, one can simplify the problem by assuming the secondary as a point--particle and solving the Teukolsky or the Regge--Wheeler--Zerilli (RWZ) equation~\cite{Aleman:2003, Khanna:2004,Burko:2007,Sundararajan:2008zm,Sundararajan:2010sr,Zenginoglu:2011zz,Fujita:2004rb,Fujita:2005kng,Mano:1996vt,throwe2010high,OSullivan:2014ywd,Drasco:2005kz} to obtain the far-field waveform. This theoretical framework is termed point--particle black hole perturbation theory (ppBHPT) and is anticipated to yield accurate waveforms for binary systems characterized by sufficiently large mass ratios. Conversely, it is expected that the accuracy of ppBHPT waveforms will diminish as we move from the extreme mass ratio limit to the comparable mass regime.

Exploring the differences between waveforms derived from NR and ppBHPT in the comparable and intermediate mass ratio regimes has long been a focal point of interest within the GW community~\cite{Price:1994pm,Lousto:2010tb,Lousto:2010qx,Nakano:2011pb}. This exploration serves a dual purpose. Firstly, it provides insights into the interaction between the NR and ppBHPT frameworks within the comparable mass regime, as well as the interplay between fully nonlinear and linear problem descriptions. Secondly, it aids in establishing the validity boundaries of the ppBHPT framework, paving the way for leveraging it to construct waveform models for BBH source characterization.

Two recent developments in extending the ppBHPT framework into the comparable mass regime are the development of a second-order self-force waveform model~\cite{Pound:2021qin,Miller:2020bft,Wardell:2021fyy} and the empirically observed efficacy of a simple model calibration technique, which we will refer to as $\alpha$-$\beta$ scaling~\cite{Islam:2022laz,Rifat:2019ltp}. In this paper, we assess the applicability of the $\alpha$-$\beta$ scaling at the extreme end of the mass ratio regime: the equal-mass binary. While previous studies explored the late-stage of equal-mass collisions in great detail within the perturbative frameworks using the close-limit approximation~\cite{Price:1994pm, Anninos:1995vf, Baker:1996bt, Nollert:1996av, Khanna:1999mh, Khanna:2000dg}, we focus on the comparison between NR and ppBHPT for the full binary evolution. Our goal is first to examine the differences between NR and ppBHPT frameworks in the equal-mass limit. We will then evaluate the efficacy of the $\alpha$-$\beta$ scaling technique well outside the presupposed expected domain of validity for ppBHPT at $q=1$. Finally, we consider possible time-dependent generalizations of the $\alpha$-$\beta$ parameters that could be used in future waveform modeling efforts.

The rest of this paper is organized as follows. In Section~\ref{sec:data}, we summarize our waveform data. Section~\ref{sec:results} outlines various methods for determining the $\alpha$-$\beta$ parameters and demonstrates their effectiveness for an equal-mass system. The limitations and implications of our study are addressed in Section~\ref{sec:discussion}.

\section{NR and ppBHPT Data at \boldmath{$q=1$}}
\label{sec:data}
We generate our $q=1$ ppBHPT waveform using the \texttt{BHPTNRSur1dq1e4}~\cite{Islam:2022laz} model. Direct simulation using our time-domain Teukolsky solver fails in the equal-mass limit due to limitations in trajectory generation for the secondary black hole that incorporates adiabatic radiative corrections). \texttt{BHPTNRSur1dq1e4} is a reduced-order surrogate model trained on ppBHPT waveform data generated from a time-domain Teukolsky equation sourced by a test particle whose adiabatic inspiral is driven by energy fluxes~\cite{Sundararajan:2007jg, Sundararajan:2008zm, Sundararajan:2010sr, Zenginoglu:2011zz}. This model is interfaced through the \texttt{BHPTNRSurrogate} package~\cite{BHPTSurrogate}, available in the \texttt{black hole perturbation} ~\texttt{Toolkit}~\cite{BHPToolkit}.  While the model has been trained for mass ratios $2.5 \leq q \leq 10^4$, surrogate models have previously demonstrated good performance when extrapolated beyond their training range~\cite{Varma:2019csw,Islam:2021mha}. In our case, extrapolation to $q=1$ will bring in error, but these errors are expected to be small~\cite{Islam:2021mha}.

We obtain NR data from a publicly accessible \texttt{SXS:BBH:1132} simulation performed by the SXS collaboration~\cite{Mroue:2013xna,Boyle:2019kee}. The NR data have $\sim50$ cycles and are $\sim$25,000$M$ long in duration, where $M:=m_1+m_2$ is the total mass of the binary. 

\section{Comparing NR and ppBHPT at \boldmath{$q=1$}}
\label{sec:results}

\subsection{Model Calibration Setup}
\label{sec:calibration}

We now investigate the differences and connections between NR and ppBHPT at $q=1$ through the lens of the $\alpha$-$\beta$ scaling. The scaling reads~\cite{Islam:2022laz}:
\begin{align} \label{eq:EMRI_rescale}
h_{\ell,m}^{\tt NR}(t^{\tt NR} ; q) \sim {\alpha_{\ell}} h_{\ell,m}^{\tt ppBHPT}\left( \beta  t^{\tt ppBHPT};q \right) \,,
\end{align}
where $h_{\ell,m}^{\tt NR}$ and $h_{\ell,m}^{\tt ppBHPT}$ represents the NR and ppBHPT waveforms, respectively, as functions of the NR time, $t^{\tt NR}$, and ppBHPT time, $t^{\tt ppBHPT}$.
These calibration parameters are found by solving an optimization problem
\begin{align} \label{eq:opt}
{\cal J} =  \min_{\alpha,\beta} \frac{\int \left| \alpha h_{2,2}^{\tt ppBHPT}\left( \beta t^{\tt ppBHPT} \right) - h_{2,2}^{\tt NR}(t^{\tt NR}) \right|^2 dt}{\int \left| h_{2,2}^{\tt NR}(t^{\tt NR}) \right|^2 dt} \,,
\end{align}
where the integral is taken over the last $\sim$5000$M$ of NR data. This allows us to determine optimal values for $\alpha$ and $\beta$. To simplify the discussion, we focus only on the waveform's $(2,2)$ mode and will use $\alpha = \alpha_2$ for brevity. Ref.~\cite{Islam:2022laz} considered subdominant modes and found that these modes follow the same pattern as the $(2,2)$ one. Furthermore, for $q=1$, the subdominant modes are very weak while odd $m$ modes are identically zero.

The gravitational wave models presented in Refs.~\cite{Rifat:2019ltp,Islam:2022laz} use the aforementioned methodology to find values for $\alpha_{\ell}(q_i)$ and $\beta(q_i)$ at mass ratios $q_i$ available from numerical relativity simulations. They are assumed to be constant (time-independent) values throughout the binary's evolution, and regression techniques are used to model their behavior over the parameter space including the $q \rightarrow \infty$ limit. The accuracy of the models developed in Refs.~\cite{Rifat:2019ltp,Islam:2022laz} relies on (i) the existence of approximately time-independent $\alpha_{\ell}(q_i)$ and $\beta(q_i)$ values and (ii) the ability to accurately build regression models for $\alpha_{\ell}(q_i)$ and $\beta(q_i)$ over a wide range of mass ratio values. This paper is mostly focused on exploring the behavior of $\alpha_{\ell}(q=1)$ and $\beta(q=1)$ as a stress test for $\alpha$-$\beta$ scaling technique. We note that in \mbox{Sections~\ref{sec:alpha_beta_comparison} and~\ref{sec:ringdown}}, we also consider a possible generalization of these parameters to time-dependent functions  $\alpha_{\ell}(t)$ and $\beta(t)$. We will continue to refer to $\alpha$ and $\beta$ as ``parameters'' even in this time-dependent context.

\subsection{Setting a Common Mass Scale}
\label{sec:mass_scale}

As discussed in Ref.~\cite{Rifat:2019ltp,Islam:2022laz}, our Teukolsky solver sets the background black hole's mass to $m_1 = 1$ and uses $m_1$ as ppBHPT's mass scale, while the corresponding NR simulation sets the total mass to $M = m_1 + m_2 = 1$ and uses $M$ as the mass scale. So before comparing waveforms, we should adjust the ppBHPT's mass scale to use the NR convention. In the ppBHPT's setup, this would be $m_1 + m_2 = 1 + 1/q$, where upon setting $q=1$, the $(2,2)$ mode should be adjusted 
according to the formula $h^{2,2}(t) \rightarrow \frac{1}{2} h^{2,2}(t/2)$ before comparing to NR. In Figure~\ref{fig:alpha_beta_example}, we show both the NR and ppBHPT waveforms after making this adjustment. While there are visual discrepancies, we find it remarkable that these waveforms are even somewhat similar, given that $q=1$ is well outside of the formal range of validity for point--particle perturbation theory.

While our $\alpha$-$\beta$ calibration procedure (Equation~\eqref{eq:EMRI_rescale}) automatically accounts for the different mass scales used, it also attempts to account for various physical effects missing in the ppBHPT setup. In keeping with conventions of previous work~\cite{Rifat:2019ltp,Islam:2022laz}, we will apply the scaling procedure described in Section~\ref{sec:calibration} to ppBHPT waveforms expressed units of $m_1$. Because the mass scale is a freely specifiable choice for expressing dimensioned quantities in terms of dimensionless ones, changing the mass scale carries no more physical meaning than, for example, expressing time in seconds instead of hours. However, we suspect the numerical values of $\alpha$ and $\beta$ do carry some physical meaning to account for missing physics (e.g., second-order self-force effects), although the interpretability of our phenomenological $\alpha$-$\beta$ scaling remains an open question. Yet, if the numerical values are to be interpreted as something physically relevant, the mass scale needs to be correctly accounted for. That is, one could imagine applying the scaling procedure described in Section~\ref{sec:calibration} to ppBHPT waveforms expressed in terms of a mass scale $M$ resulting in a different set of parameters $\{ \hat{\alpha}, \hat{\beta}\}$, which are related by $\alpha = \frac{1}{1+1/q} \hat{\alpha}$ and $\beta = \frac{1}{1+1/q} \hat{\beta}$. For our $q=1$ system, $\hat{\alpha} = 2 \alpha$ and $\hat{\beta} = 2 \beta$ would be the relevant values to compute and where, for example, if $\hat{\alpha} = \hat{\beta} = 1$, then NR and ppBHPT waveforms would be physically identical.

\subsection{Effectiveness of the $\alpha$-$\beta$ Scaling at $q=1$}
\label{sec:alpha_beta_q1}
\begin{figure}[H]

\begin{adjustwidth}{-\extralength}{0cm}
\centering
\includegraphics[width=14.5 cm]{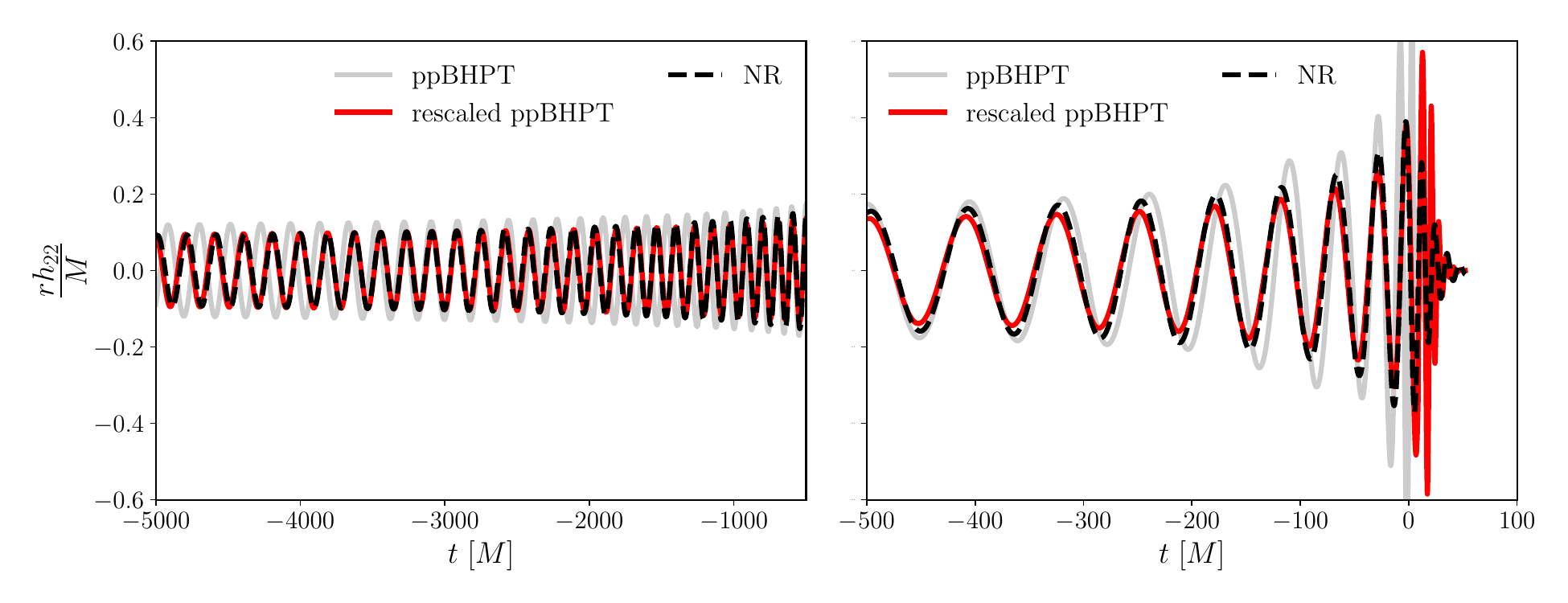}
\end{adjustwidth}
\caption{A non-spinning $q=1$ waveform for the $(2,2)$ mode from numerical relativity (black dashed line), ppBHPT (grey line), and the calibrated ppBHPT (red lines). The left panel shows the waveforms for the early inspiral, while the right panel focuses on the merger-ringdown stage. All waveforms have the same mass scale of $M=m_1 + m_2$. }
\label{fig:alpha_beta_example}
\end{figure}

We first consider whether a simple scaling (Equation~\eqref{eq:EMRI_rescale}) exists for equal-mass binaries. We find that Equation~\eqref{eq:opt} is minimized at $\alpha=0.402599$ and $\beta=0.455557$
where ${\cal J}=0.038$.
In Figure~\ref{fig:alpha_beta_example}, we show both the NR and rescaled ppBHPT waveforms. Additionally, for comparison, we show the ppBHPT waveform after only accounting for the mass scale differences as described in Section~\ref{sec:mass_scale}.

It is worth noting that while the calibrated ppBHPT and NR waveforms exhibit excellent agreement during the inspiral phase, noticeable differences become apparent in the plunge--ringdown stage. These differences can be attributed to a combination of various simplifying assumptions in the ppBHPT framework; a non-inclusive list includes: (i) the final remnant black hole's mass and spin value is assumed to be the same as the primary black hole's initial mass and spin, (ii) the orbital plunge model is expected to be inaccurate in the comparable mass limit, (iii) finite size effects are likely to be critically important for an equal-mass binary close to merger, and (iv) non-linear, strong gravity effects (e.g., second-order metric perturbations) may become important. For accurate modeling, these effects must be accounted for~\cite{Islam:2023aec,Islam:2023mob}. However, as extremely accurate GW models in the comparable mass regime already exist, even a sufficiently well-calibrated ppBHPT-based waveform model in its current form is unlikely to deliver competitive models at $q=1$.

One source of error in our calibrated ppBHPT waveform (see Equation \eqref{eq:EMRI_rescale}) could be the breakdown of the assumption that $\alpha$ and $\beta$ are time-independent. In the following subsection, we investigate the possible time dependence of these parameters in detail.
\subsection{Exploring the Time Dependence of $\alpha$-$\beta$ Parameters from Different Methods}
\label{sec:alpha_beta_comparison}

\begin{figure}[H]
\includegraphics[width=13.5 cm]{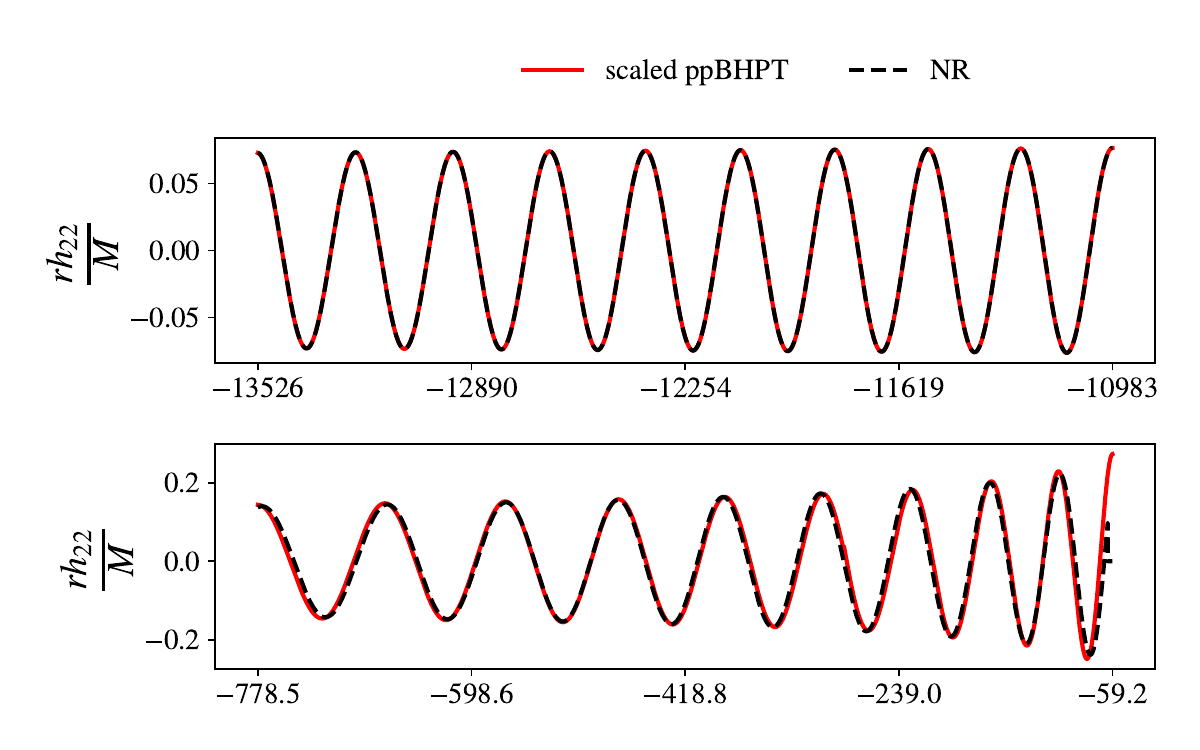}
\caption{The waveform's $(2,2)$ mode from NR (black dashed line) and scaled ppBHPT (red solid line) calibrated over a short 5-cycle window of data: $t\in[-13526M,-10983M]$ and $t\in[-778.5M,-59.2M]$. This demonstrates that the $\alpha$ and $\beta$ parameters are nearly time-independent over in the early inspiral portion, an assumption that breaks down in the late inspiral. Further details can be found in Section~\ref{sec:alpha_beta_comparison}.}
\label{fig:5cycle}
\end{figure}

Figure~\ref{fig:alpha_beta_example} clearly suggests that, for equal-mass binaries, both $\alpha$ and $\beta$ must have some temporal dependence. To illustrate this issue further, we compute time-varying $\alpha$ and $\beta$ values using two different methods described in Ref.~\cite{Islam:2023jak}.
The first method seeks to find pointwise values of $\alpha$ and $\beta$, such that the amplitude and phasing of the ppBHPT waveform match NR at each NR wave peak. We denote these estimates as $\alpha_{\rm peak}$ and $\beta_{\rm peak}$. For the second method, we select the NR and ppBHPT data encompassing the final 35 cycles before the merger and segment them into seven consecutive 5-cycle windows. Each 5-cycle window is composed of 10 peaks. It is important to note that these waveform windows will have different time durations between NR and ppBHPT data but will maintain an identical number of peaks. Within each of these time windows, we minimize $\cal{J}$ from  Equation~\eqref{eq:opt} yielding local estimates of $\alpha$ and $\beta$ applicable in each time window. We refer to these estimates as $\alpha_{\rm 5cycles}$ and $\beta_{\rm 5cycles}$. Using the second method, 
Figure~\ref{fig:5cycle} shows both the rescaled ppBHPT waveform (in red solid lines) and NR data (in black dashed lines) over the first and last 5-cycle time windows. Notice that in the last 5-cycle window, covering $t=-778.5M$ to $t=-59.2M$, visible differences emerge between the rescaled ppBHPT waveform and NR, indicating the assumption of a time-independent $\alpha$-$\beta$ is no longer valid even within this restricted signal duration.
\vspace{-12pt}

\begin{figure}[H]
\includegraphics[width=11 cm]{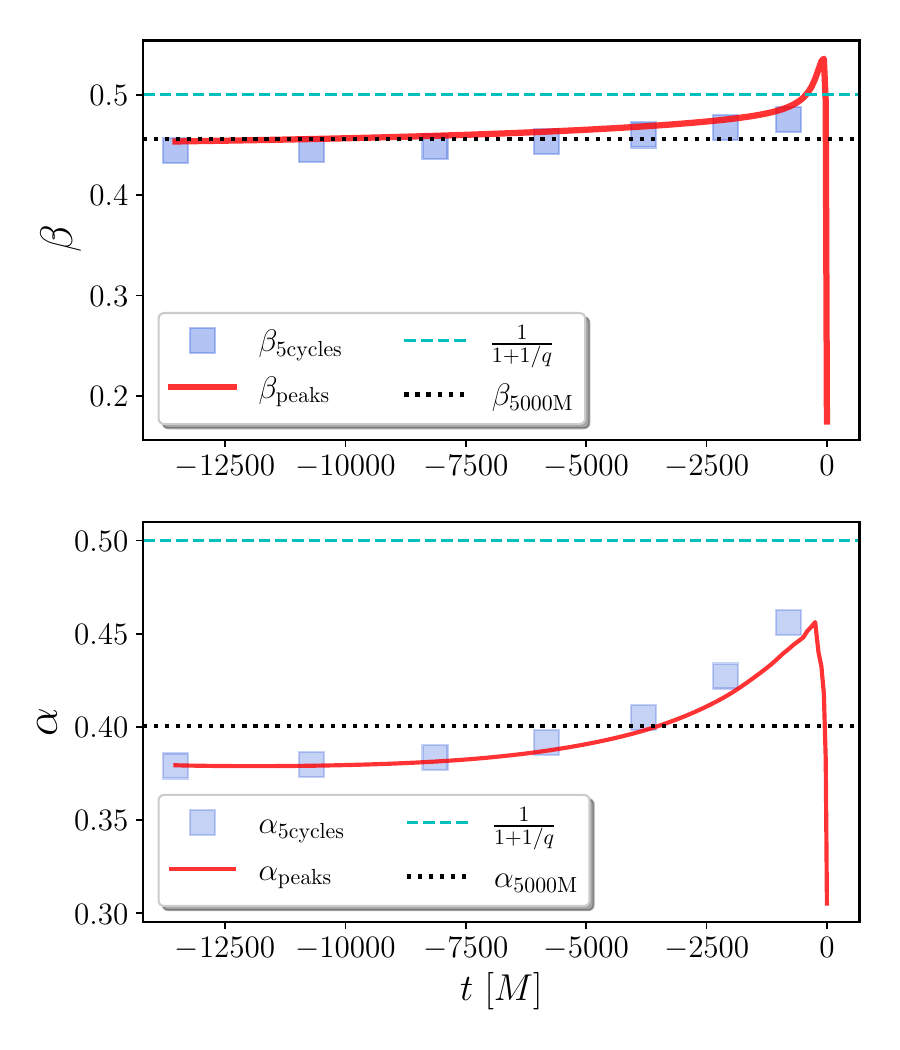}
\caption{Calibration parameters, $\alpha$ and $\beta$, for a $q=1$ BBH system obtained from different approaches outlined in Section~\ref{sec:alpha_beta_comparison}; these are labeled \{``5cycles'', ``5000M'', ``peaks''\}. For comparison, we also show the scaling of $\frac{1}{1+1/q} = 0.5$ required to change the ppBHPT mass scale to match the NR one. If $\alpha=\beta=0.5$, then the NR and ppBHPT waveform's $(2,2)$ modes are identical. More details can be found in Section~\ref{sec:alpha_beta_comparison}.}
\label{fig:alpha_beta_vs_time}
\end{figure}

In Figure~\ref{fig:alpha_beta_vs_time}, we compare calibration parameter values obtained using all three methods described above. For comparison, we also show the scaling value of $\frac{1}{1+1/q}=0.5$ required to make the mass scale of ppBHPT and NR waveforms consistent, as discussed in Section~\ref{sec:mass_scale}. We find that $\alpha$ and $\beta$ values obtained from these different approaches differ from $0.5$, demonstrating that the calibration parameters are necessary to account for physical effects missing in the ppBHPT model.
We also notice that $\alpha_{\rm peak}$ and  $\beta_{\rm peak}$ remain relatively constant in the early inspiral and then quickly rise near the merger. This explains why the time-independent $\alpha$-$\beta$ scaling used in previous models~\cite{Islam:2022laz,Rifat:2019ltp} effectively matches the inspiral waveform but does not perform as well in the late inspiral--merger--ringdown phase. 

\begin{figure}[H]
\includegraphics[width=11 cm]{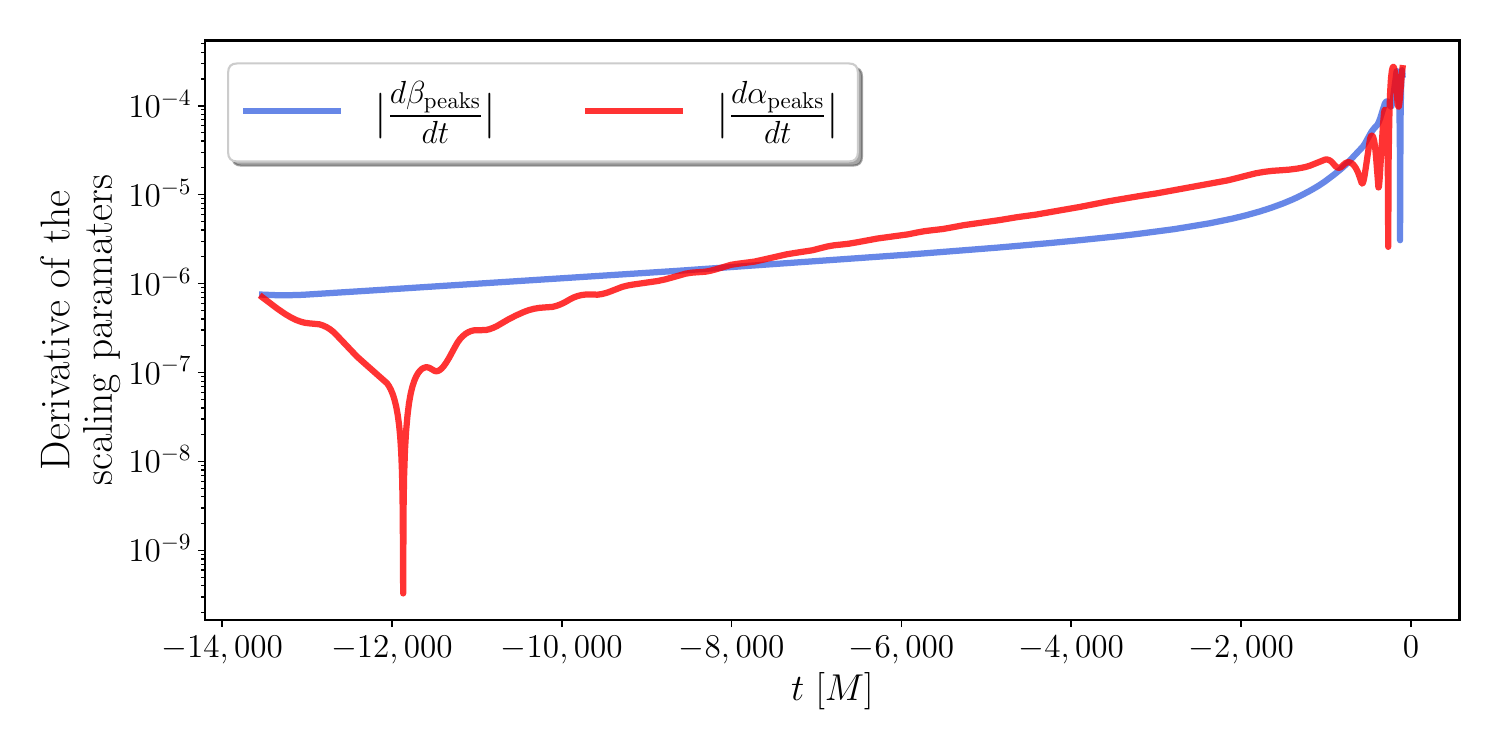}
\caption{Time derivative of the scaling parameters $\alpha_{\rm peak}$ and $\beta_{\rm peak}$ obtained locally using the "peaks" method summarized in Section~\ref{sec:alpha_beta_comparison}. While the derivatives are not zero, they remain small and increase in the late-inspiral stage. This explains why the $\alpha$-$\beta$ calibration technique, which assumes time-independent values for $\alpha$ and $\beta$, works particularly well in the inspiral. Further details can be found in Section~\ref{sec:alpha_beta_comparison}.}
\label{fig:alpha_beta_dot}
\end{figure}

To emphasize the temporal evolution of the scaling parameters, we show derivatives of both $\alpha_{\rm peak}$ and $\beta_{\rm peak}$ in Figure~\ref{fig:alpha_beta_dot}. While subtle, the temporal evolution of $\alpha$ and $\beta$ introduces errors in scaled ppBHPT waveforms, as illustrated in Figure~\ref{fig:alpha_beta_example}. If we can incorporate this time dependence into the scaled ppBHPT waveforms, it has the potential to significantly enhance our modeling accuracy. However, achieving this improvement necessitates a systematic investigation into the temporal evolution of $\alpha$ and $\beta$ over the parameter space of binary systems. We leave this endeavor for future exploration.

\unskip

\unskip

\subsection{Understanding $\alpha$-$\beta$ Values at the Ringdown}
\label{sec:ringdown}
\begin{figure}[H]
\includegraphics[width=7.5 cm]{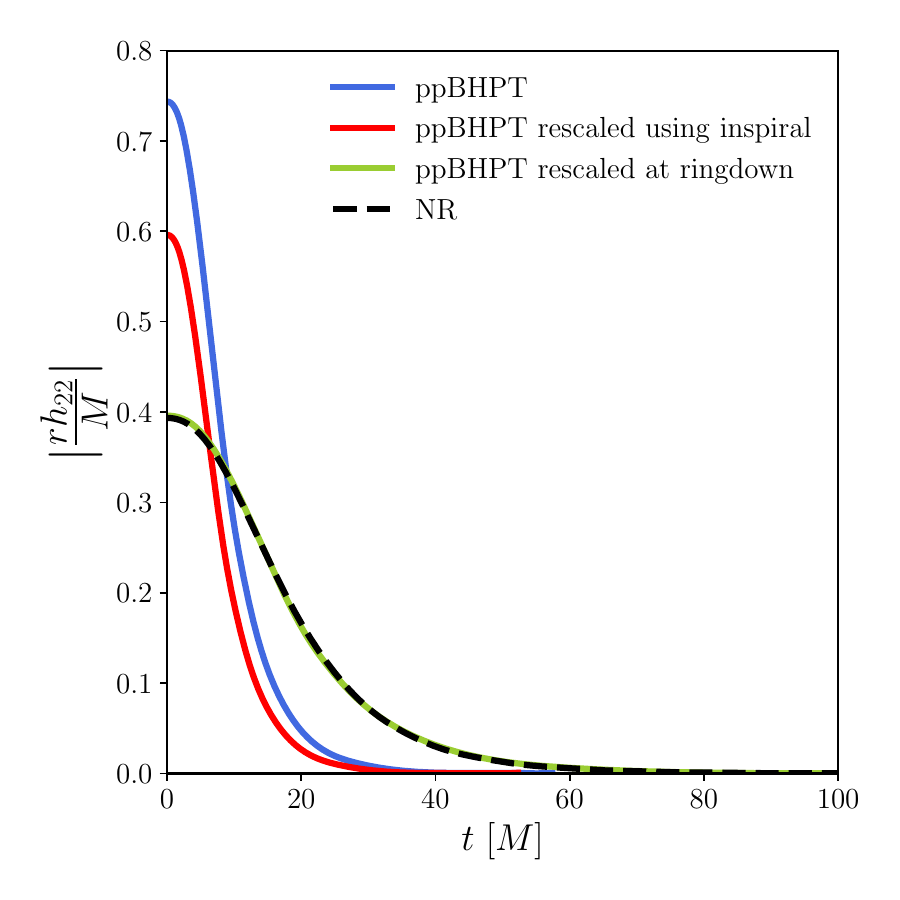}
\caption{The ringdown waveform's $(2,2)$ mode from NR (black dashed line), ppBHPT waveform (blue solid line), ppBHPT waveform rescaled using only $\alpha$ and $\beta$ parameters determined from the inspiral data (red solid line), and ppBHPT waveform rescaled using  $\alpha$ and $\beta$ parameters determined from the ringdown data (green solid line). While the $\alpha$ and $\beta$ parameters determined from the inspiral are unsuitable for the ringdown, a different set of parameter values---constant over the final $100$M of the signal---can again deliver an accurate calibration. Further details can be found in Section~\ref{sec:ringdown}.}
\label{fig:rd}
\end{figure}

The final piece of our investigation focuses on the ringdown phase of the waveform where, as shown in Figure~\ref{fig:alpha_beta_example}, the most significant disparities between NR and rescaled ppBHPT waveforms are observed. This prompts the question of whether alternative $\alpha$ and $\beta$ values can be found. We minimize Equation~\eqref{eq:opt} over the final $100M$ of the signal, finding optimal values $\beta=1.063$ and $\alpha=0.266$, which is significantly different from the inspiral-based values of $\beta=0.455557$ and $\alpha=0.402599$. In Figure~\ref{fig:rd}, we show both the NR and rescaled ppBHPT ringdown signals using scaling parameter values that arise from different methods. We see that while the $\alpha$ and $\beta$ parameters determined from the inspiral are not appropriate for the ringdown, a different set of parameter values---constant over the final $100$M of the signal---can again deliver an accurate calibration.
\vspace{-6pt}

\unskip

\section{Discussion and Conclusions}
\label{sec:discussion}

In this paper, we compared NR and ppBHPT waveforms for an equal-mass binary black hole system. Given that $q=1$ is well outside of the formal range of validity for the point--particle perturbation theory, this serves as a stress test for ppBHPT. We investigate the relationship between these ostensibly disparate frameworks through a phenomenological $\alpha$-$\beta$ scaling approach previously introduced as a simple yet effective method to calibrate ppBHPT waveforms to NR. This approach applied to the waveform's (2,2) mode only requires two time-independent parameters to scale the overall amplitude and time coordinate of the ppBHPT waveforms. 

Our findings indicate that the $\alpha$-$\beta$ scaling~\cite{Islam:2022laz,Rifat:2019ltp} (as shown in Equation \eqref{eq:EMRI_rescale}) performs reasonably well for equal-mass binaries, except during the late inspiral--merger--ringdown phase. When compared against NR waveforms of about $5000M$ in duration, the scaled ppBHPT waveform yields an $L_2$-norm error of $0.038$. As shown here, this error is largely due to the time-independent assumption of $\alpha$-$\beta$. For larger mass ratios, this assumption works better, and highly accurate models can be built~\cite{Islam:2022laz}. For comparable mass ratios, this breakdown is likely associated with multiple limiting assumptions of ppBHPT, including the importance of finite size effects and the remnant black hole's mass and spin values. Furthermore, by generalizing our calibration parameters to be time-dependent, we illustrate that these parameters (i) have minimal temporal variation in the early inspiral and (ii) exhibit greater temporal variation as the binary system is about $\approx$5 orbits before the merger. We also show that ringdown-specific scaling values---that are time-independent over the last $100M$ of the signal---can be found. A comprehensive exploration of the merger-ringdown data for various mass ratios and spins will be important for future waveform calibration endeavors using this method. Despite these challenges, our investigation yields hopefully insightful observations regarding NR and ppBHPT waveforms under the constraints of equal-mass binaries. This work offers interesting insights into the effectiveness of the $\alpha$-$\beta$ scaling in the equal-mass limit as well as its limitations. For example, introducing a certain amount of time dependence for the calibration parameters $\alpha$ and $\beta$ may allow for better-calibrated  ppBHPT-based models, such as \texttt{BHPTNRSur1dq1e4}. We leave this for future work.

One potential limitation of our analysis is that the ppBHPT waveform is generated using \texttt{BHPTNRSur1dq1e4} (a surrogate model) outside its training range. Yet direct simulation of a $q=1$ system with the ppBHPT framework was not possible as our time-domain Teukolsky solver~\cite{Sundararajan:2007jg, Sundararajan:2008zm, Sundararajan:2010sr, Zenginoglu:2011zz} breaks down at the equal-mass limit (more specifically, the trajectory generation for the secondary black hole that incorporates adiabatic radiative corrections fails in the equal-mass limit). While surrogate models can be extrapolated a bit beyond their training range~\cite{Varma:2019csw,Islam:2021mha}, this will bring in additional errors that might compromise studies that require very high accuracy.

\vspace{6pt} 

\authorcontributions{Conceptualization, T.I.; methodology, T.I., G.K., S.E.F.; software, T.I., G.K., S.E.F.; investigation, T.I.; resources, T.I., G.K., S.E.F.; data curation, T.I., S.E.F.; writing---original draft preparation, T.I.; writing---review and editing, T.I., G.K., S.E.F.; visualization, T.I.; funding acquisition, G.K., S.E.F.
}

\funding{This information is already provided in the acknowledgments section. Should the journal wish to move this information here, it would read ``The authors acknowledge the support of NSF grants PHY-2106755, PHY-2307236 (G.K) and DMS-1912716, DMS-2309609 (T.I and G.K).  SEF acknowledges support from NSF grants PHY-2110496, DMS-2309609, and UMass Dartmouth's Marine and Undersea Technology (MUST) research program funded by the Office of Naval Research (ONR) under grant no. N00014-23-1–2141. Simulations were performed on CARNiE at the Center for Scientific Computing and Data Science Research (CSCDR) of UMassD, which is supported by  ONR/DURIP grant no.\ N00014181255 and the UMass-URI UNITY supercomputer supported by the Massachusetts Green High-Performance Computing Center (MGHPCC).}

\dataavailability{No new data were created or analyzed in this study. Data sharing is not applicable to this article.}

\acknowledgments{Submitted to a themed issue in honor of Jorge Pullin on his 60th anniversary. The authors acknowledge the support of NSF grants PHY-2106755, PHY-2307236 (G.K) and DMS-1912716, DMS-2309609 (T.I and G.K).  SEF acknowledges support from NSF grants PHY-2110496, DMS-2309609, and UMass Dartmouth's Marine and Undersea Technology (MUST) research program funded by the Office of Naval Research (ONR) under grant no. N00014-23-1–2141. Simulations were performed on CARNiE at the Center for Scientific Computing and Data Science Research (CSCDR) of UMassD, which is supported by  ONR/DURIP grant no.\ N00014181255 and the UMass-URI UNITY supercomputer supported by the Massachusetts Green High Performance Computing Center (MGHPCC).}

\conflictsofinterest{The authors declare no conflict of interest} 

\bibliography{References}
\PublishersNote{}
\end{document}